\documentclass[longbibliography]{appolb}
\usepackage{graphicx}
\usepackage{enumitem}
\DeclareGraphicsExtensions{.pdf,.png,.jpg}
\usepackage{amsmath} 
\usepackage{amssymb}    
\usepackage{latexsym}
\usepackage{amsfonts}
\usepackage{mathrsfs}
\usepackage{color}
\usepackage{bbold}
\usepackage[colorlinks=true, urlcolor=blue,citecolor=blue,linkcolor=blue]{hyperref}

\newcommand{\bra}[1]{\langle #1|}
\newcommand{\ket}[1]{|#1\rangle}

\newcommand{\abs}[1]{\left|#1\right|}
\newcommand{\avg}[1]{\left<#1\right>}

\newcommand{\cre}[1]{a^{\dagger}_{#1}}
\newcommand{\ani}[1]{a_{#1}}

\begin{document}

\title{Entanglement conditions involving intensity correlations of optical fields: the case of multi-port interferometry}

\author{Junghee Ryu$^{1,2}$, Marcin Marciniak$^{2}$, Marcin Wie\'{s}niak$^{2,3}$, Dagomir Kaszlikowski$^{1,4}$, and Marek \.{Z}ukowski$^{2,5}$
\address{$^1$Centre for Quantum Technologies, National University of Singapore, 3 Science Drive 2, 117543 Singapore, Singapore}
\address{$^2$Institute of Theoretical Physics and Astrophysics, Faculty of Mathematics, Physics and Informatics, University of Gda\'{n}sk, 80-308 Gda\'{n}sk, Poland}
\address{$^3$Institute of Informatics, Faculty of Mathematics, Physics and Informatics, University of Gda\'{n}sk, 80-308 Gda\'{n}sk, Poland}
\address{$^4$Department of Physics, National University of Singapore, 2 Science Drive 3, 117542 Singapore, Singapore}
\address{$^5$Faculty of Physics, University of Vienna, Boltzmanngasse 5, A-1090 Vienna, Austria}
}

\maketitle

\begin{abstract}
Normalized quantum Stokes operators introduced in [Phys. Rev. A {\bf 95}, 042113 (2017)] enable one to better observe non-classical correlations of entangled states  of optical fields with undefined photon numbers. For a given run of an experiment the new quantum Stokes operators are defined by the differences of the measured intensities (or photon numbers) at the exits of a polarizer divided by their sum. It is this ratio that is to be averaged, and not the numerator and the denominator separately, as it is in the conventional approach. The new approach allows to construct more robust entanglement indicators against photon-loss noise, which can detect entangled optical states in situations in which witnesses using standard Stokes operators fail. Here we show an extension of this approach beyond phenomena linked with polarization. We discuss EPR-like experiments involving correlations produced by optical beams in a multi-mode bright squeezed vacuum state. EPR-inspired entanglement conditions for all prime numbers of modes are presented. The conditions are much more  resistant to noise due to photon loss than similar ones which employ standard Glauber-like intensity, correlations. 
\end{abstract}

\section{Introduction}
A new approach to the analysis of polarization correlations was introduced in \cite{ZUKOWSKI17}. It involves a non-conventional definition of Stokes parameters of quantum optical fields. The ``textbook'' quantum Stokes parameters involve averaged differences of the measured intensities of light exiting polarization analyzers, measured in three complementary arrangements (horizontal-vertical, diagonal-anti-diagonal and right-left-handed circular polarizations), and the total average intensity. If one assumes that the measured intensities are proportional to photon numbers, then the textbook Stokes parameters read
\begin{equation}
\avg{\hat{\mathbb{\Sigma}}_{i}} = \Tr[(\cre{i}\ani{i} - \cre{i \perp} \ani{i \perp}) \varrho],
\end{equation}
where $\ani{i}$ is the annihilation operator for photons of polarization $i$, $\ani{i \perp}$ for the orthogonal polarization, and the index $i$ denotes the three complementary arrangements. One also has $\hat{\mathbb{\Sigma}}_0=\cre{i}\ani{i} + \cre{i \perp} \ani{i \perp}=\hat{N}_{\rm tot}^{a}$.

However, this approach to Stokes parameters faces problems when the intensities of quantum optical fields are fluctuating. This is so for example in the case for the optical fields generated multi-mode  parametric down-conversion,  especially for  higher  pump  powers. This is due to the fact that the traditional Stokes parameters depend on  intensity fluctuations.  Experimental runs involving higher measured intensities contribute more to their values. However, polarization of light is an intensity independent phenomenon.

It turns out that one can revise the concept of quantum Stokes observables, and remove this conceptual difficulty. The new approach allows one to derive much more effective indicators of non-classicality for quantum optical fields with undefined, fluctuating, intensities. To this end, Ref.~\cite{ZUKOWSKI17} introduces normalized Stokes operators. They are based on the ratios of the numbers of photons registered by one of the two detectors, to the number of photons counted by both detectors {\it for a given run} (not averages). In new formalism the redefined, normalized Stokes parameters read
\begin{equation}
\avg{\hat{\mathcal{S}}_i} = \Tr \left[ \Pi \, \frac{\cre{i} \ani{i} -\cre{i \perp} \ani{i \perp}}{\hat{N}_{\rm tot}^{a}}\, \Pi \,\varrho \right],
\label{EQ:NSTOKES}
\end{equation}
where $\Pi$ denotes a projection written $\hat{I} - \ket{\Omega}\bra{\Omega}$ with the identity operator $\hat{I}$ and the vacuum state $\ket{\Omega}$ for modes $i$ and $i_{\perp}$, which simply removes the vacuum terms in the state $\varrho$, and makes the operator in the numerator well defined.
Note that in case of vacuum events, that is, measurements showing $N_{\rm tot} = 0$, we have no contribution to  the redefined Stokes parameters (and neither to the standard ones). The $\hat{\mathbb{\Sigma}}_0$ parameter is replaced by $\hat{\mathcal{S}}_0=\Pi$, and simply gives the probability of a non-vacuum event.



The redefined Stokes operators allow a derivation of new quantum optical Bell inequalities~\cite{ZUKOWSKI16}, which are an improvement of standard ones introduced for general optical intensities in \cite{REID86}. The new Bell inequalities are based only on the standard Bell assumptions of realism, locality and free will, and nothing more. They  do not require any specific additional ``reasonable" assumptions on the form of the hidden variable theories, which  are necessary to derive  the standard ones of Ref.~\cite{REID86}. Most importantly the new ``theoretical-loophole" free Bell inequalities constructed for the redefined Stokes operators can be violated by classes of entangled optical fields for which the standard ones are not violated. E.g., this is the case for four-mode (bright) squeezed vacuum (BSV), generated in type-II parametric down conversion for stronger pump powers. Note that this is a typical example of a laboratory situation, where often parametric down conversion is used to get entanglement effects. For stronger laser pumping   the squeezed vacuum  has many contributing Fock components  with different photon numbers - we have exactly the aforementioned situation of undefined photon numbers. 

The new Stokes operators also allow to re-formulate any known entanglement conditions~\cite{ZUKOWSKI17}. In case of the four-mode, BSV states, Ref.~\cite{ZUKOWSKI17} reformulates an entanglement indicator of Ref.~\cite{SIMON03}, which reads
\begin{equation}
\sum_{i} \avg{\left({\mathbb{\Sigma}}_{i}^{a} +{\mathbb{\Sigma}}_{i}^{b}\right)^2 }_{\rm sep} \geq 2 \avg{\hat{N}_{\rm tot}^{a} + \hat{N}_{\rm tot}^{b}}_{\rm sep},
\label{EQ:SIMONSEP}
\end{equation}
where $\avg{\cdots}_{\rm sep}$ denotes an average over a separable state. Here, $a$ and $b$ denote two beams (defined by propagation directions). The lower bound is proportional to the averaged total photon numbers, i.e., the sum of averaged total intensities of  beams $a$ and $b$. The idea of the  indicator is based on the Einstein-Podolsky-Rosen (EPR)-like condition which is satisfied by the rotationally invariant bright squeezed vacuum: for this state one has a zero value for the left hand side expression in~(\ref{EQ:SIMONSEP}). Such a state is essentially a super-singlet with undefined photon numbers, which for identical polarization settings at two separated observation stations, always gives anti-correlated results.

In Ref.~\cite{ZUKOWSKI17}, the photon number operators, like e.g., $\cre{i}\ani{i}$, were in the left hand side in~(\ref{EQ:SIMONSEP}) replaced by  the rates, respectively $\Pi \, \frac{\cre{i}\ani{i}}{\hat{N}_{\rm tot}^{a}}\, \Pi$. For the rotationally invariant squeezed vacuum, we still have the EPR condition. However one can show that for a separable state one has
\begin{equation}
\sum_{i} \avg{ \left(\hat{\mathcal{S}}_{i}^{a} + \hat{\mathcal{S}}_{i}^{b}\right)^2 }_{\rm sep}
 \geq 2 \left( \avg{\Pi^{a} \frac{1}{\hat{N}^{a}_{\rm tot}} \Pi^{a}}_{\rm sep} + \avg{\Pi^{b} \frac{1}{\hat{N}^{b}_{\rm tot}} \Pi^{b}}_{\rm sep} \right).
\label{EQ:STOKESSEP}
\end{equation}
The new condition turns out to be more robust with respect to photon losses, modeled in~\cite{ZUKOWSKI17} by inefficient detectors.

Generally, re-formulating the separability conditions with new Stokes operators enable a much better entanglement detection in case of BSV states. New condition is more robust against photon losses (or non-perfect detection efficiency). The previous condition~(\ref{EQ:SIMONSEP}) fails to detect the entanglement of  the BSV state for the detection efficiency lower than $1/3$ (see~\cite{SIMON03}), while the new one (\ref{EQ:STOKESSEP}) still works for lower efficiencies. The actual threshold of the efficiency is a decreasing function of gain parameter which depends on the pump power, interaction time of the pump laser, and the coupling~\cite{ZUKOWSKI17}. Furthermore, the approach allows one to map any entanglement conditions for two qubits in such a way, that we obtain conditions for polarization of quantum optical fields. Thus, we can now construct a plethora of new entanglement conditions for correlations of quantum light. The approach also allows a better visibility of some non-classical phenomena, see~\cite{ROSOLEK17}, in which the working example is the Hong-Ou-Mandel effect~\cite{HONG87}, under strong pumping condition.

Here, we extend this approach beyond polarization effects. In Ref.~\cite{RYUJOPT17} one can find an approach based of rates, rather than intensities, for two three-mode  optical beams (the example of a state used there was a six-mode bright squeezed vacuum). The entanglement criteria are again  inspired by properties of EPR correlations, which are impossible for any separable state. Here we present a generalization of this approach to two $d$-mode beams. For technical reasons, which will be discussed below, we assume that  $d$ is a prime. After finishing this manuscript we  have found a condition which works for all $d$, which are powers of a prime number. This will be reported elsewhere.

\section{
EPR-like optical experiments involving pairs of multi-port beam splitters}
We shall consider here a class of entanglement experiments analyzed in Ref.~\cite{ZUKOWSKI97}, allowing for EPR-like correlations~\cite{ZEILINGER93}, which involve measurement basis transformations defined by two local multi-port interferometers (beam-splitter). Such interferometers were first studied in~\cite{WALKER87}, however not in the context of entanglement. 

It is known that $d$-input-port-$d$-output-port interferometers can perform any finite dimensional unitary $d \times d$ transformations of single particle states described by a $d$ dimensional Hilbert space, see Ref.~\cite{RECK}. Such interferometers consist of interconnected beam splitters and phase shifters. The  same blueprint for  multi-port devices allows for optical modes coupling resulting in a unitary relation way between the input and output modes. For instance, consider a single photon  prepared in input mode $i=0,1,...,d-1$ denoted by $\ket{\phi_i^{\rm in}}$. We assume throughout that the input and output modes are fully distinguishable. The multi-port action can be described by the unitary transformation which gives as output states  $|\phi_k^{\rm out}\rangle=\sum_i U_{ki}\ket{\phi_i^{\rm in}}$, where  $U$ is unitary matrix. The equivalent mode transformation  for the  second quantized description can be written down in terms of  photon creation operators (related with the `in' and `out' modes, or beams, $i=0,1,...,d-1$): 
\begin{equation}
a_k^{\dagger {\rm out}}=\sum_i U_{ki}{a_i^{\dagger {\rm in}}}.
\label{EQ1}
\end{equation}
Beacause of the assumed distinguishability among the `in' modes and among the `out' modes, we assume that  $[a^{{\rm in}}_i, a_j^{\dagger {\rm in}} ]=\delta_{ij}$, and  $[a^{{\rm in}}_i, a^{{\rm in}}_j]=0$, and similar relations for the `out' modes.

With the advent of integrated optics, which allows for  stable complicated interferometers, two  multi-port  experiments, such as the ones suggested in~\cite{ZUKOWSKI97}, are becoming feasible. Recently, the work~\cite{Schaeff15} tested exactly such configurations. The schemes discussed here involve parametric down-conversion for higher pump powers, in the case of which we have superpositions of multi-pair emissions. Thus new phenomena can be expected, which can be both decremental or beneficial for possible quantum communication experiments. At least one should check to what extent the features of two-photon correlations related with entanglement and the EPR paradox,  are still present in the case of stronger fields.

We consider quantum optical states produced by multi-mode emissions in the parametric down-conversion  process \cite{ZUKOWSKI97, PAN}. Due to the phase matching conditions, the emissions from a parametric down-conversion source are directionally correlated. For example, type-I parametric down-conversion process generates the pairs of photons of the same frequency with emission directions which form a cone. One can register coincidences into pairs of ``conjugated" directions along the cone which lay in the same plane as the pump field, for details see \cite{PAN}. The directions of such pairs satisfy the phase matching condition. We can select an arbitrary number of such pairs of the directions, and collect their radiation.
In such a case the description of the crystal-field interaction leading to the process can be given by  an interaction Hamiltonian of the form:
\begin{eqnarray}
H= {\rm i} \gamma \sum_{i=0}^{d-1} a_{i}^{\dagger} b_{i}^{\dagger} + h.c.,
\label{EQ:H_INT}
\end{eqnarray}
where $a_{i}^{\dagger}$ and $b_{i}^{\dagger}$ are the creation operators of $i$-th conjugate signal-idler mode pair, and $\gamma$ is a coupling constant proportional to the pumping power. The mode operators $a_i$ and $b_i$ refer to two conjugate directions. 

Notice that the Hamiltonian in~(\ref{EQ:H_INT}) can be transformed into the following form:
\begin{eqnarray}
H= {\rm i} \gamma \sum_{k=0}^{d-1}  a_{k}^{\dagger \rm out} b_{k}^{\dagger \rm out} + h.c.,
\label{EQ:OUTMODE_H}
\end{eqnarray}
where $a_k^{\dagger \rm out}=\sum_j U_{kj}{a_j^\dagger}$, and 
$b_k^{\dagger \rm out}=\sum_i U^*_{ki}{b_i^\dagger}$ with a $d\times d$ unitary matrix $U$. Due to this symmetry of the Hamiltonian $H$, one has the perfect EPR correlations for the emitted photons. EPR correlations must occur for at lease two complementary operational situations, thus to see that the symmetry implies, it is enough to consider a transformation $U$ which leads to a complementary mode basis. This is when the observation bases at location $a$ and $b$ are allowing to measure respectively $\cre{i}\ani{i}$ and $b_{j}^{\dagger} b_{j}$, where $i,j = 0,1,\dots,d-1$.




For prime $d$ pairs of Schmidt modes, the emitted photon pairs are prepared in the following entangled state:
\begin{eqnarray}
\ket{{\rm BSV}}=\frac{1}{\cosh^{d} \Gamma} \sum_{n=0}^{\infty} \sqrt{\frac{(n+d-1)!}{n!(d-1)!}} \tanh^n\Gamma \ket{\psi^{n}},
\label{EQ:BSV}
\end{eqnarray}
where
\begin{eqnarray}
\ket{\psi^n}=\sqrt{\frac{n!(d-1)!}{(n+d-1)!}}\sum_{p_1+\cdots+p_d=n} \ket{p_1}_{a_1} \cdots \ket{p_d}_{a_d} \ket{p_1}_{b_1} \cdots \ket{p_d}_{b_d}.
\label{EQ:BSV_2}
\end{eqnarray}
The sum is taken over all combinations of nonnegative integers $p_i$. The parameter $\Gamma$ describes the gain and is dependent on coupling constant $\gamma$ and the interaction time (which can be put as equal to the length of the non-linear crystal, along the propagation direction of the laser field, divided by the speed of light).

The local measurement devices which we consider consist of an unbiased or ``symmetric" \cite{ZUKOWSKI97},  multi-port beam-splitter  and $d$ detectors in its output ports, which are by assumption capable to resolve photon numbers. An unbiased multi-port beam-splitter is defined as a $d$-input and $d$-output interferometric device which realizes a mode transformation defined by a unitary matrix $U$ which links two unbiased orthonormal bases in the $d$ dimensional Hilbert space. In the case of such transformations a single photon entering through a single port can be detected at any of the output ports with  the same probability of $1/d$. A simple analytic formula holds for the values of  matrix elements for such unitary transformations only for  $d$ which is prime. This is the reason why we concentrate here on this case. It is known that $d+1$
mutually unbiased bases exist for all Hilbert spaces of dimensions which are powers of primes~\cite{Wootters,Ivanovic}. In other cases, including the simplest one $d=6$, the number of possible unbiased bases is still an open question. Finally, note that if two bases are unbiased this means that they describe two perfectly complementary operational measurement situations.

In~\cite{RYUJOPT17}, the separability conditions for the three-output case are studied. We here shall derive the separability conditions for arbitrary $d$ which is prime. We shall also use a different approach.
\begin{figure}[t]
\center
\includegraphics[width=11cm]{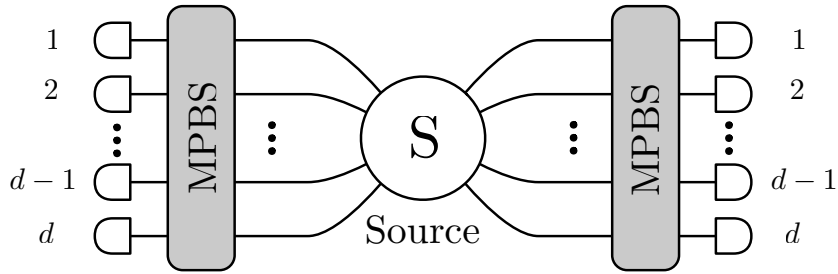}
\caption{Schematic diagram of the experiment. The local measurement stations  consist of a $d$-input-$d$-output multi-port beam-splitter (denoted by MPBS) and detectors. The interaction in the source $S$, given by the Hamiltonian (\ref{EQ:H_INT}), generates a $d$-mode bright squeezed vacuum state (\ref{EQ:BSV}). This entangled state leads to  perfect EPR correlations between the local conjugate modes.}
\label{FIG:state_d3}
\end{figure}


\section{Entanglement indicators for prime $d$}
Consider $d+1$ unitary transformations which lead to the unbiased (complementary) bases in a $d$-dimensional Hilbert space. It is known that when the dimension $d$ of a Hilbert space is an integer power of a prime number, the number of mutually unbiased bases is given by $d+1$~\cite{Wootters,Ivanovic}. We put $U(d)=\hat{I}$, while the others, indexed with $m=0,1, \dots, d-1$, have matrix elements which lead to the following transformations of the bases \cite{Wootters,Ivanovic}: 
\begin{eqnarray}
U(m)_{js} = \frac{1}{\sqrt{d}} \omega^{js +ms^2},
\end{eqnarray}
where $\omega=\exp(2 \pi {\rm i} /d)$. With such transformations one can relate a multi-port beam-splitter which couples the creation operators of input beams, $a_s^\dagger$, with the output ones, $a_{j}^{\dagger} (m)$ in the following way: 
\begin{eqnarray}
a_{j}^{\dagger} (m) = \frac{1}{\sqrt{d}} \sum_{s=0}^{d-1} \omega^{js +ms^2} a_{s}^{\dagger},
\label{EQ:CRE_D}
\end{eqnarray}
and we define $a_{j}^{\dagger} (m=d) =  a_{j}^{\dagger}$. The photon number operator of $j$th exit mode of an $U(m)$ multi-port reads $\hat{n}_{j}(m) = \cre{j}(m) \ani{j}(m)$. Note that for beams $b$ we have a conjugate unitary relation, compare relation (\ref{EQ:OUTMODE_H}) and its discussion.

For the (bright) squeezed vacuum, $\ket{\rm BSV}$, for $d$ pairs of modes ($d$ is a prime), its perfect correlations give us the following relation:
\begin{eqnarray}
\avg{\sum_{m=0}^{d} \sum_{j=1}^{d} \left[ \hat{n}_{j}^{A} (m) - \hat{n}_{j}^{B} (m)\right]^2}_{\rm BSV} =0,
\label{BSV-EPR}
\end{eqnarray}
where indices $A$ and $B$ denote operators for Alice and Bob, respectively.
The aim now is to find the lower bound of such an expression for any separable state. As separable states are convex combinations of product ones, 
it is enough to find 
the minimum of the expression \eqref{BSV-EPR} for a product state  $\varrho_{A} \otimes \varrho_{B}$ instead of BSV, where both $\varrho_A$ and $\varrho_B$ are pure states. We have
\begin{eqnarray}
&&\avg{\sum_{m=0}^{d} \sum_{j=1}^{d} \left[ \hat{n}_{j}^{A} (m) - \hat{n}_{j}^{B} (m)\right]^2}_{\varrho_{A} \otimes \varrho_{B}} \nonumber \\
&=& \sum_{m,j}  \avg{\hat{n}_{j}^{A} (m)^2}_{\varrho_{A}} + \sum_{m,j} \avg{\hat{n}_{j}^{B} (m)^2}_{\varrho_{B}} -2 \sum_{m,j} \avg{\hat{n}_{j}^{A} (m)}_{\varrho_{A}} \avg{\hat{n}_{j}^{B} (m)}_{\varrho_B} \nonumber \\
&\geq& \sum_{m,j} \avg{\hat{n}_{j}^{A} (m)^2}_{\varrho_{A}} + \sum_{m,j} \avg{\hat{n}_{j}^{B} (m)^2}_{\varrho_{B}} \nonumber \\
&&{}-2 \left(\sum_{m,j} \avg{\hat{n}_{j}^{A} (m)}_{\varrho_{A}}^{2}\right)^{1/2} \left( \sum_{m,j}\avg{\hat{n}_{j}^{B} (m)}_{\varrho_B}^{2}\right)^{1/2}.
\label{EQ:ENT}
\end{eqnarray}


Let us first consider one of the factors of the term in the last line  of the inequality \eqref{EQ:ENT}  (we drop below the index numbering the measuring stations $A, B$):
\begin{eqnarray} \label{EQ:EXP_SQRT-1}
&&\sum_{m=0}^{d} \sum_{j=1}^{d} \avg{\hat{n}_{j} (m)}^2 = \sum_{j=1}^{d} \avg{\hat{n}_{j}}^2 + \sum_{m=0}^{d-1} \sum_{j=1}^{d} \avg{\hat{n}_{j} (m)}^2.
\end{eqnarray}
Let us now consider the second term in the above equation, involving summation over  $m$ only up to $d-1$. Using~(\ref{EQ:CRE_D}), we get  the following:
\begin{eqnarray}
\label{EQ:EXP_SQRT}
&& \sum_{m=0}^{d-1} \sum_{j=1}^{d} \avg{\hat{n}_{j} (m)}^2\nonumber \\ 
&=& \frac{1}{d^2}\sum_{m=0}^{d-1} \sum_{j=1}^{d} \avg{\sum_{s,t} \omega^{j (s-t) + m(s^2 - t^2)} a_{s}^{\dagger} a_{t}}^2 \nonumber \\
&=&  \frac{1}{d^2} \sum_{m,j}  \sum_{s_1, t_1} \sum_{s_2, t_2} \omega^{j (s_1 - t_1 + s_2 - t_2) + m(s_{1}^{2} - t_{1}^{2} +s_{2}^{2} -t_{2}^{2})} \avg{a_{s_1}^{\dagger} a_{t_1}} \avg{a_{s_2}^{\dagger} a_{t_2}} \nonumber \\
&=& \frac{1}{d^2} \sum_{s_1, t_1} \sum_{s_2, t_2}\avg{a_{s_1}^{\dagger} a_{t_1}} \avg{a_{s_2}^{\dagger} a_{t_2}} \sum_{j} \omega^{j (s_1 - t_1 + s_2 - t_2)} \sum_{m}  \omega^{m(s_{1}^{2} - t_{1}^{2} +s_{2}^{2} -t_{2}^{2})}  \nonumber \\
&=&  \frac{1}{d^2} \left\{ \sum_{s_1 = t_1} \sum_{s_2=t_2}+ \sum_{s_1 = t_1} \sum_{s_2\neq t_2}+ \sum_{s_1\neq t_1} \sum_{s_2=t_2} + \sum_{s_1 \neq t_1} \sum_{s_2\neq t_2}\right\} \nonumber \\
&& \times \avg{a_{s_1}^{\dagger} a_{t_1}} \avg{a_{s_2}^{\dagger} a_{t_2}} \sum_{j} \omega^{j (s_1 - t_1 + s_2 - t_2)} \sum_{m}  \omega^{m(s_{1}^{2} - t_{1}^{2} +s_{2}^{2} -t_{2}^{2})}  \nonumber \\
&=&  \frac{1}{d^2} \left[ d^2\sum_{s_1, s_2} \avg{\hat{n}_{s_1}} \avg{\hat{n}_{s_2}}+ \sum_{s_1 \neq t_1} \sum_{s_2 \neq t_2}\avg{a_{s_1}^{\dagger} a_{t_1}}\avg{a_{s_2}^{\dagger} a_{t_2}} \right. \nonumber \\
&&\times  \left.  \sum_{j} \omega^{j (s_1 - t_1 + s_2 - t_2)} \sum_{m}  \omega^{m(s_{1}^{2} - t_{1}^{2} +s_{2}^{2} -t_{2}^{2})}  \right].\label{mid}
\end{eqnarray}
The last equality follows from the fact that $s_1-t_1+s_2-t_2\neq 0$ when $s_1=t_1$, $s_2\neq t_2$ or $s_1\neq t_1$, $s_2=t_2$.

Now, we will show that for the case $s_1\neq t_1$, $s_2\neq t_2$, the expression $\sum_{j} \omega^{j (s_1 - t_1 + s_2 - t_2)} \sum_{m}  \omega^{m(s_{1}^{2} - t_{1}^{2} +s_{2}^{2} -t_{2}^{2})}$ is nonzero if and only if $s_1=t_2$ and $s_2=t_1$. To this end, observe that the nonzero value occurs only if
\begin{eqnarray}
s_1-t_1+s_2-t_2&=&0, \label{eq1} \\
s_1^2-t_1^2+s_2^2-t_2^2 &=& 0. \label{eq2}
\end{eqnarray}
It follows from \eqref{eq1} that $s_2-t_2=-(s_1-t_1)\neq 0$. Thus, the second equation implies
\begin{equation}
s_1+t_1-s_2-t_2=0. \label{eq3}
\end{equation}
Equations \eqref{eq1} and \eqref{eq3} lead to the conditions $s_1=t_2$ and $s_2=t_1$. Finally, we observe that if these conditions are satisfied, then  
\begin{eqnarray}
\sum_{j}\omega^{j (s_1 - t_1 + s_2 - t_2)} \sum_{m}\omega^{m (s_{1}^{2} - t_{1}^{2} + s_{2}^{2} - t_{2}^{2})} = d^2. \nonumber
\end{eqnarray}
Thus, the formula (\ref{EQ:EXP_SQRT-1}) is reduced to
\begin{eqnarray}
\label{EQ:EXP_SQRT_2}
&& \sum_{j} \avg{\hat{n}_j}^2 + \sum_{s_1, s_2} \avg{\hat{n}_{s_1}} \avg{\hat{n}_{s_2}} + \frac{1}{d^2} \sum_{s_1 \neq t_1} d^2 \avg{a_{s_1}^{\dagger} a_{t_1}} \avg{a_{t_1}^{\dagger} a_{s_1}}\nonumber \\
&=& \sum_{j} \avg{\hat{n}_j}^2 + \avg{\hat{N}}^2 + \sum_{s_1 \neq t_1} \left<{a_{s_1} \psi} |a_{t_1} \psi \right> \left<{a_{t_1} \psi} |a_{s_1} \psi \right>\nonumber \\
&\leq& \sum_{j} \avg{\hat{n}_j}^2 + \avg{\hat{N}}^2 + \sum_{s_1 \neq t_1} || a_{s_1} \psi ||^2 || a_{t_1} \psi ||^2 \nonumber \\
&=& \sum_{j} \avg{\hat{n}_j}^2 + \avg{\hat{N}}^2 + \sum_{s_1 \neq t_1} \avg{\hat{n}_{s_1}} \avg{\hat{n}_{t_1}} = 2 \avg{\hat{N}}^2,
\end{eqnarray}
where $\ket{\psi}$ is a normalized vector which determines the pure state $\varrho_X$ ($X=A,B$), $\ket{a_{t_1} \psi}=a_{t_1}\ket{ \psi}$, and $ || a_{t_1} \psi ||$ is its norm. 
Finally, we have 
\begin{equation}
\sum_{m=0}^d\sum_{j=1}^d \avg{\hat{n}_{j} (m)}^2 \leq 2\avg{\hat{N}}^2.
\end{equation}

In case of the first two terms in (\ref{EQ:ENT}), we have 
\begin{eqnarray}
\label{EQ:SINGLE}
&&\sum_{j} \hat{n}_{j}^{2} + \frac{1}{d^2} \sum_{m=0}^{d-1} \sum_{j=1}^{d} \sum_{s_1, t_1, s_2, t_2} \omega^{j (s_1 - t_1 + s_2 - t_2) + m(s_{1}^{2} - t_{1}^{2} +s_{2}^{2} -t_{2}^{2})} a_{s_1}^{\dagger} a_{t_1} a_{s_2}^{\dagger} a_{t_2} \nonumber \\
&=& \sum_{j} \hat{n}_{j}^{2} + \frac{1}{d^2} \sum_{s_1, t_1, s_2, t_2} \left(\sum_{j} \omega^{j (\cdots)} \right) \left(\sum_{m} \omega^{m (\cdots)} \right) a_{s_1}^{\dagger} a_{t_1} a_{s_2}^{\dagger} a_{t_2} \nonumber \\
&=& \sum_{j} \hat{n}_{j}^{2} + \sum_{s_1, s_2} \hat{n}_{s_1} \hat{n}_{s_2} + \sum_{s_1 \neq t_1} a_{s_1}^{\dagger} a_{t_1} a_{t_1}^{\dagger} a_{s_1}\nonumber \\
&=& \sum_{j} \hat{n}_{j}^{2} + \sum_{s_1, s_2} \hat{n}_{s_1} \hat{n}_{s_2} + \sum_{s_1 \neq t_1} a_{s_1}^{\dagger} a_{s_1} (a_{t_1}^{\dagger} a_{t_1}+1) \nonumber \\
&=& \sum_{j} \hat{n}_{j}^{2} + \sum_{s_1, s_2} \hat{n}_{s_1} \hat{n}_{s_2} + \sum_{s_1 \neq t_1} \hat{n}_{s_1} \hat{n}_{t_1} + (d-1) \sum_{s_1} \hat{n}_{s_1} \nonumber \\
&=& \hat{N}^2 + (d-1) \hat{N} + \hat{N}^2 = 2 \hat{N}^2 + (d-1)\hat{N}
\end{eqnarray}
To obtain this, we used again the observation from the paragraph following \eqref{mid}.
With the help of the results of (\ref{EQ:EXP_SQRT_2}) and (\ref{EQ:SINGLE}), finally one derives the separability condition (\ref{EQ:ENT}) in the form of
\begin{eqnarray}
&&\avg{\sum_{m=0}^{d} \sum_{j=1}^{d} \left[ \hat{n}_{j}^{A} (m) - \hat{n}_{j}^{B} (m)\right]^2}_{\varrho_{A} \otimes \varrho_{B}} \nonumber \\
&=&2 \avg{\hat{N}^{A}}^2 + (d-1)\avg{\hat{N}^{A}} + 2 \avg{\hat{N}^{B}}^2 +(d-1)\avg{\hat{N}^{B}} - 4 \avg{\hat{N}^{A}} \avg{\hat{N}^{B}}\nonumber \\
&\geq& (d-1) \left(\avg{\hat{N}^A} + \avg{\hat{N}^B} \right).
\label{EQ:INT_CON}
\end{eqnarray}
To get an analog conditions for the rates, one has to retrace the above derivation, replacing the number operators by the rates $\hat{r}_{j}(m) \equiv \Pi \, \hat{n}_{j} (m) \frac{1}{\hat{N}}\,\Pi $. The condition for the rates reads:
\begin{eqnarray}
&&\avg{\sum_{m=0}^{d} \sum_{j=1}^{d} {[\hat{r}_{j}^{A} (m) - \hat{r}_{j}^{B} (m)]^2}}_{\rm sep} \nonumber \\
&&\geq (d-1) \left( \avg{\Pi^{A} \frac{1}{\hat{N}^{A}} \Pi^{A}}_{\rm sep} + \avg{\Pi^{B} \frac{1}{\hat{N}^{B}} \Pi^{B}}_{\rm sep} \right).
\label{EQ:FINALSEP_NUM}
\end{eqnarray}

For $d=3$, this condition is equivalent to the one derived in~\cite{RYUJOPT17}, and thus as shown there it is more robust with respect to noise related with photon losses. Our numerical studies show that the condition (\ref{EQ:FINALSEP_NUM}) outperforms (\ref{EQ:INT_CON}) also for higher $d$'s. 

\section{Complementarity relations}
As a by-product, a kind of complementarity relations for arbitrary prime $d$ follow from the relations (\ref{EQ:EXP_SQRT_2}), which read
$\sum_{m}\sum_{j}\avg{ \hat{n}_{j} (m)}^2 \leq 2\avg {\hat{N}}^2$, and their analog for the rates $\hat{r}_{j}(m) $ reads 
\begin{equation} 
\label{COMPL-2}
\sum_{m=0}^{d} \sum_{j=1}^{d} \avg{ \hat{r}_{j} (m)}^2 \leq 2.
\end{equation}
Thus, if e.g., $\avg{\hat{r}_1(m)}=1 $, then $\sum_{m}\sum_{j \neq 1}\avg { \hat{r}_{j} (m)}^2 \leq 1.$ As a matter of fact,  $\avg{\hat{r}_1(m)}=1$ implies that the state in question describes all photons exiting via beam $1$ (or exit $1$) of the multi-port beam-splitter related with the complementary situation $m$, and also no vacuum component in the state. This implies that in such a case for all the other complementary situations $m' \neq m$, and each $j$th exit, one has $\avg{\hat{r}_j(m')}=1/d$. Thus the relation (\ref{COMPL-2}) is a form of the usual property of mutually unbiased bases, for complementary interferometers and arbitrary optical fields.

\section{Summary and closing remarks}
For  $2\times d$-mode (where $d$ is a prime) quantum optical fields of undefined intensities, we formulate series of separability criteria based on observed intensities~(\ref{EQ:INT_CON}), and observed rates~(\ref{EQ:FINALSEP_NUM}). As an example, we consider a $d$-mode bright squeezed vacuum state. Such optical states have a EPR-like correlations of numbers of photons registered  in conjugated modes and therefore they violate the conditions~(\ref{EQ:INT_CON}) and~(\ref{EQ:FINALSEP_NUM}). With the help of multi-port integrated optics beam-splitter techniques, observation of such correlation becomes feasible.  As the critical efficiencies for  our entanglement conditions are quite moderate, for $\ket{\rm BSV}$ as our numerical studies show they are below $1/(d+1)$, the conditions can find application in analysis of experiments.

The condition~(\ref{EQ:FINALSEP_NUM}) is capable to detect entanglement,  in situations in which the one based on intensities, given by~(\ref{EQ:INT_CON}), fails.
All this concurs with our conjecture  that the correlation functions involving rates rather than intensities  can  become a useful tool in quantum optics. We expect that one can find benefits by using the rates in various cases, e.g., quantum steering, and etc., see our forthcoming manuscripts.

The results can be generalized to all $d$ for which $d+1$ mutually unbiased bases are known to exits. See a forthcoming paper.

{\em Acknowledgments.} The works was initiated as part of the BRISQ2 FP7-ICT EU grant no. 308803. MZ acknowledges  COPERNICUS DFG/FNP award-grant (2014-2018). The team of authors was additionally supported by TEAM project of FNP. JR acknowledges the National Research Foundation, Prime Minister’s Office, Singapore and the Ministry of Education, Singapore under the Research Centres of Excellence programme, and DZ acknowledges National Research Foundation and Ministry of Education in Singapore. MW acknowledges UMO-2015/19/B/ST-2/01999.

\end{document}